\documentclass[12pt]{iopart}

\usepackage[]{caption}
\usepackage{epsfig} 
\usepackage{iopams} 
\usepackage{amssymb}
\usepackage{amsfonts}

\begin{document}

\title{Pattern formation in colloidal mixtures under external
  driving fields}


\author{J. Dzubiella and H. L{\"o}wen} 
\address{Institut f{\"u}r Theoretische Physik II,
Heinrich-Heine-Universit\"at D{\"u}sseldorf,
Universit\"atsstra{\ss}e 1, D-40225 D\"usseldorf, Germany}

\eads{\mailto{joachim@thphy.uni-duesseldorf.de}}

\begin{abstract}
The influence of an external field acting differently on the two
constituents of a binary colloidal mixture performing Brownian
dynamics is investigated by computer
simulations and a simple theory. In our model,  
one half of the particles ($A$-particles)
are pulled by an external force ${\vec F}^{(A)}$
 while the other half of them ($B$-particles) are
 pulled  by an external force ${\vec F}^{(B)}$.
If ${\vec F}^{(A)}$ and ${\vec F}^{(B)}$  are parallel
and the field-free state is a mixed fluid, 
previous simulations (J. Dzubiella et al,
Phys. Rev. E {\bf 65} 021402  (2002)) have shown a
nonequilibrium pattern formation  
involving lanes of $A$ or $B$ particles only
which are sliding against each other in the direction
of the external forces. In this paper,
we generalize the situation both to non-parallel external forces and to 
 field-free crystalline states. For non-parallel
forces, lane formation is also observed but with an orientation
{\it tilted} with respect to the external forces. If the field-free state
is crystalline, a continuous increase of the parallel external forces
yields a novel {\it reentrant freezing} behavior: the crystal
first melts mechanically via the external force  and then recrystallizes into
demixed crystalline lanes sliding against each other. 
    
\end{abstract}
\pacs{05.70.Ln, 61.20.Ja, 82.70.Dd, 64.70.Dv}

\submitto{J. Phys.: Condens. Matter (Les Houches meeting: "Liquid
  state theory: from white dwarfs to colloids" in honor to J. P. Hansen).}

\date{\today }

\maketitle

\section{Introduction}

When brought into nonequilibrium, physical systems may
spontaneously exhibit many different kinds of pattern formation
(for recent reviews see Refs.\ \cite{pattern,Saarloos}) which are much richer
than the traditional phase transitions in equilibrium systems.
While the latter are by now well-understood by microscopic theories
and simulations \cite{Allen,review1,Hartmutr,HansenMcDonald},
full microscopic theories operating on a particle-resolved level
for nonequilibrium situations still
represent a major challenge. In this paper we study a ``microscopic''
system designed to model binary colloidal suspensions in an external
field which is an off-lattice version \cite{Ziareview} of a diffusive system
in an external driving field.  Colloidal dispersions indeed represent excellent model
systems which can be brought into nonequilibrium \cite{ramaswamy1,colloid_non}
via external fields in a controlled way \cite{top}
and can be studied experimentally with a real-space resolution of  an interparticle
distance \cite{Murray1996}. Another complementary realization of a 
driven diffusive off-lattice systems is  pedestrian motion
in pedestrian zones \cite{Helbing,Helbing_neu,schadschneider}.

In our model, we consider an equimolar binary mixture of so-called $A$ and $B$ particles.
The particles are identical as far as their mutual interaction is concerned which we model
via a Yukawa pair potential having charged colloidal dispersions in mind.
The $A$ and $B$ particles, however, respond differently to the external field applied:
$A$ particles feel an external force ${\vec F}^{(A)}$ while $B$-particles
are driven  by a different force ${\vec F}^{(B)}$.
Completely overdamped Brownian dynamics (with hydrodynamic interactions neglected)
are assumed \cite{Hoffmann1,Hoffmann3} for the colloidal motion.
The case of parallel forces where ${\vec F}^{(A)}=-{\vec F}^{(B)}$ was investigated
recently by computer simulation by us and G. Hoffmann \cite{jo} via nonequilibrium
Brownian dynamics computer simulations in the case where the field-free thermodynamic
equilibrium state was a mixed fluid. As a result, above a critical strength of the external
force, the system exhibits a transition towards lane formation. The lanes comprise
bundles of particles of the same kind ($A$ or $B$) and are parallel to the driving field. 
This nonequilibrium phase separation \cite{rothman:1994}results from a slip-stream effect caused
or transported by the interparticle interactions.
The critical force can be theoretically estimated by setting the external force
to a typical interaction force resulting from a  pairwise potential $V(r)$
between the particles. Experimental evidence for such an instability
has been accumulated in sedimentation dynamics of bidisperse
suspensions \cite{weiland,batchelor,nasreldin:1999,yan:1993}.

In this paper, we generalize the set-up previously studied in Ref.\  \cite{jo}
into the following two directions: First, we study the case where the two external 
forces ${\vec F}^{(A)}$ and ${\vec F}^{(B)}$ are not parallel. Second, we 
study - for parallel forces - the case where the equilibrium field-free state is 
crystalline. For non-parallel forces, lane formation does also occur. The direction
of the lanes, however, is tilted with respect to the driving fields. In more
details lanes are directed along the difference vector ${\vec F}^{(B)}-{\vec F}^{(A)}$
of the two external forces. For a randomly occupied crystal, on the other hand, a two-stage
transition shows up: first, above a threshold,  the external fields melts the solid 
mechanically.
Upon increasing the external field strength further, a reentrant freezing 
transition is discovered. The resulting crystalline structure
involves completely  demixed $A$- and $B$ solids sliding against each other
similar to the fluid lane formation.

The paper is organized as follows: In section \ref{model}, we define the
model used and describe briefly our simulation technique. Results for
non-parallel external forces are presented in section \ref{nonparallel}.
The solid-fluid-solid reentrant behavior is discussed in section \ref{reentrant}.
Conclusion are given in section \ref{conclusions}. In particular, we
discuss a possible verification of our predictions in experiments.

\section{The model}
\label{model}

In our model \cite{jo}, we consider a symmetric binary colloidal mixture comprising 
$2N$ Brownian colloidal particles in $d=2$ spatial dimensions. 
Half of them are particles of type $A$, the other half is of type
$B$ with  partial number densities  $\rho_{A} = \rho_{B} =
\rho /2$. The colloidal suspension  is held at fixed temperature $T$
via the  bath of microscopic solvent particles.
Two colloidal particles are interacting via an effective Yukawa
pair potential
\begin{equation}
V(r) = V_0\,\sigma\exp\left[ -\kappa (r-\sigma)/\sigma \right]/r,
\label{interaction}
\end{equation}
where $r$ is the center-to-center separation,
$V_0$ is an energy scale and $\sigma$ is the particle diameter as a length scale.
This is a valid model for charge-stabilized suspensions confined to two dimensions
\cite{Hone,Loewen92,Loehle}.

The dynamics of the colloids is overdamped
Brownian motion. The friction constant $\xi = 3\pi \eta \sigma$ 
(with $\eta$ denoting the shear 
viscosity of the solvent) is assumed to be the same for both $A$
and $B$
particles. 
The constant external force acting on the $i$th particle, $\vec
F_{i}$, is different for the both constituents of the binary mixture. 
It is $\vec F_{i}={\vec F}^{(A)}$ for $A$ particles and $\vec F_{i}={\vec
  F}^{(B)}$ for $B$ particles.

The stochastic
Langevin equations for the colloidal trajectories ${\vec r}_i(t)$ 
$(i=1,...,2N)$ read as
\begin{equation}
\xi \frac{{d{\vec r}_i}}{dt} = -{\vec \nabla}_{{\vec r}_i} \sum_{j\not= i} V
(\mid {\vec r}_i -{\vec r}_j \mid ) + {\vec F}_i  + {\vec
  F}_i^{(\rm R)}(t). 
\label{langevin}
\end{equation}
The right-hand-side includes all forces acting onto
the colloidal particles, namely the force resulting from
inter-particle interactions, the external constant force,
and the random forces ${\vec F}_i^{(\rm R)}$ describing 
 the collisions of the solvent molecules with the $i$th
 colloidal particle. The latter
are Gaussian random numbers with zero mean, $\overline {{\vec
    F}_i^{(\rm R)}}=0$,
and variance 
\begin{equation}
\overline{({\vec F}_i^{(\rm R)})_{\alpha }(t)({\vec F}_j^{(\rm R)})_{\beta }(t')}={{2k_BT}{\xi}}
\delta_{\alpha\beta} \delta_{ij}\delta(t-t').
\label{variance}
\end{equation}
The subscripts $\alpha$ and $\beta$ stand for the two 
Cartesian components and $k_{B}T$ is the thermal energy.
In the absence of an external field, the model reduces to a two-dimensional  
Brownian Yukawa fluid in equilibrium which has 
been extensively investigated as far as structural and 
dynamical equilibrium correlations and freezing transitions
are concerned 
\cite{Loewen92,Loehle,Naidoo}.

We solve the Langevin equations of motion by Brownian
dynamics simulations 
\cite{Hoffmann1,Hoffmann2,Roux} using a finite time-step and the technique of Ermak 
\cite{Allen,Ermak}. We  observed the system running into a steady-state after a transient relaxation.
We put $N=250$ $A$ and $N=250$ $B$ particles 
into a square cell of length $\ell$  with 
periodic boundary conditions. The total colloidal number density is
$\rho = 2N/\ell^2$.
The typical size of the time-step was $0.003 \tau_{\rm
  B}$, where  $\tau_{B}=\xi\sigma^{2}/V_{0}$ is a suitable  Brownian timescale.
We simulated typically $2\times 10^4$ time steps which corresponds 
to a  simulation time of $60\tau_{B}$. After 
an initial relaxation period
of  $20\tau_{B}$, statistics was  gathered in the steady state.

\section{Non-parallel external forces}
\label{nonparallel}

\subsection{General argument for tilted lane formation}
Let us first recapitulate what is know for {\it parallel} external
forces ${\vec F}^{(A)}$ and ${\vec F}^{(B)}$: in Ref.\ \cite{jo},
it was shown that lane formation occurs 
involving either $A$ or $B$ particles which are sliding against
each other in the field direction. In the lane involving $A$ particles
only, all these particle are drifting with an global
velocity ${\vec F}^{(A)}/\xi$ while  opposite regions which
involve $B$ particles are streaming with the overall
velocity ${\vec F}^{(B)}/\xi$. By subtracting the overall velocity
using a Galilei transformation one readily sees that within the completely
separated lanes, equilibrium Boltzmann statistics is realized.
The system just separates into two different equilibrium states
which are drifting relative to each other. Physically, the formation of lanes
is generated by collisions of $A$ against $B$ particles pushed by the external force
which dynamically separates $A$ and $B$ particles until completely demixed
lanes are formed. A similar lane formation for sheared granular matter
was found via molecular dynamics simulation in \cite{santra:1996}. The
formation of lanes is a  sharp first-order  non-equilibrium 
phase transition occurring if the external field difference
$| {\vec F}^{(A)}-{\vec F}^{(B)}|$ exceeds a critical value.
A Galilei transformation also proves that only the relative velocity
of $A$ and $B$ regions is relevant. 
Hence, without loss of generality, it is sufficient to
study the special case ${\vec F}^{(A)}=-{\vec F}^{(B)}$.

\begin{figure}
\begin{center}
    \epsfig{file=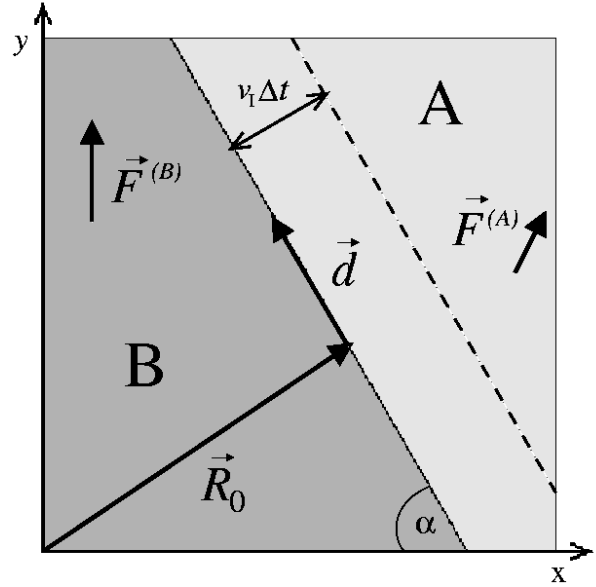, width=8cm, angle=0}
    \caption{One-dimensional interface separating a region containing
      $A$ particles (light gray) and $B$ particles only (dark
      gray). The direction of the interface is $\vec d$. The vector
      $\vec R_{0}$ points to the interface. The new interface after a
      time $\Delta t$ at distance $v_{\rm I}\Delta t$ from the
      original one is shown as a dashed line.}
  \label{fig1}
\end{center}
\end{figure}

For {\it non-parallel} external forces, the collisions between $A$ and $B$ particles
are not any longer central and the phase separated structure will be different, in general.
In order to get insight into the location of an interface between two
completely demixed regions involving $A$ and $B$ particles only, we first do
a simple continuum argument: consider a (one-dimensional) interface between an $A$
and $B$ region with a direction described by a two-dimensional vector ${\vec d}$,
see Fig. \ref{fig1}. The full interface position can be parameterized
by a set of vectors

\begin{eqnarray}
{\vec r} (s) = {\vec R}_{0} + s {\vec d},
\label{interface1}
\end{eqnarray}
where $s$ is a real parameter and $\vec R
_{0}$ is a vector describing a point on the interface.
After a time $\Delta t$, the $A$ particles have moved on average
a distance ${\vec F}^{(A)}\Delta t /\xi$, while the $B$ particles 
have been displaced by a distance ${\vec F}^{(B)}\Delta t /\xi$.
Neglecting any collisions, the $A$ particles near the interface will move
towards  a new interface which is described by the set of vectors
\begin{eqnarray}
{\vec r}^{(A)} (s) = {\vec R}_{0} + s {\vec d} +{\vec F}^{(A)}\Delta t /\xi,
\label{interface2}
\end{eqnarray}
while the $B$ particles near the interface will drift to
\begin{eqnarray}
{\vec r}^{(B)} (s') = {\vec R}_{0} + s' {\vec d} +{\vec F}^{(B)}\Delta t /\xi
\label{interface3}
\end{eqnarray}
with another real parameter $s'$.
The interface can only be stable if these two interfaces coincide. In case
they do not, there is either empty space which will be filled
by neighboring particles or $A$ and $B$ particles will collide which will destroy
the interface as well. Hence, the stability criterion is that for any $s$ there
is an $s'$ such that
${\vec r}^{(A)} (s)={\vec r}^{(B)} (s')$ which
simply yields the condition
\begin{eqnarray}
 {\vec d} = {{\Delta t} \over {\xi (s-s')}} [{\vec F}^{(B)} - {\vec F}^{(A)}].
\label{interface4}
\end{eqnarray}
This implies that for an interface to be stable, its direction
has to be collinear with the force difference
\begin{eqnarray}
\Delta \vec F=\vec F^{(B)}-\vec F^{(A)}.
\label{diff}
\end{eqnarray}
Hence, the angle $\alpha$ describing the interface orientation (see
Fig. \ref{fig1}) is
\begin{eqnarray}
\alpha=\arcsin\frac{\Delta\vec F\cdot\vec F^{(B)}}{|\Delta\vec  F||\vec F^{(B)}|}.
\end{eqnarray}
Clearly, contrarily to the case of parallel forces, the interface position will
move in space. The interface velocity $v_{I}$ normal to its position 
can be calculated as
\begin{eqnarray}
v_{I} = {{| {\vec F}^{(A)} \times {\vec F}^{(B)} |}\over {\xi | \Delta \vec F |}}.
\label{velocity}
\end{eqnarray}
Obviously, the same argument can be repeated with exchanged roles of $A$ and $B$
showing that stable parallel lanes with the direction $\Delta \vec F$ 
are expected with move with the interface velocity $v_{I}$  
given by Eq.\  (\ref{velocity}). Furthermore, 
the same argument applied for a small $\Delta t$ shows that {\it a curved interface} is
unstable such that a stable interface has to be straight.

Let us finally discuss two special cases: first returning
to parallel forces, indeed the interface direction is parallel to the field
direction, the angle $\alpha$ is $\pi/2$, and the interface velocity
vanishes as follows directly  
from Eq.\ (\ref{velocity}).
Second the case of perpendicular forces deserves some particular attention.
Here the angle $\alpha$ is
\begin{eqnarray}
\alpha=\arctan{ \frac{| {\vec F}^{(B)} |} {| {\vec F}^{(A)} |}  }
\label{angle}
\end{eqnarray}
and the interface velocity can be expressed as 
\begin{eqnarray}
v_{I} = {{| {\vec F}^{(A)} | | {\vec F}^{(B)} |}\over {\xi | \Delta \vec F |}}.
\label{velocity2}
\end{eqnarray}

Obviously, this general argument is only a necessary condition for a stable $AB$ interface.
An alternative is a mixed situation with no interface at all driven by entropy.
In analogy to the parallel case we anticipate that a critical strength
of $\Delta \vec F$ is needed to build stable lanes. This will now
be investigated by computer simulation and simple theory in more detail.

\subsection{Computer simulation results}
\begin{figure}
\begin{center}
    \epsfig{file=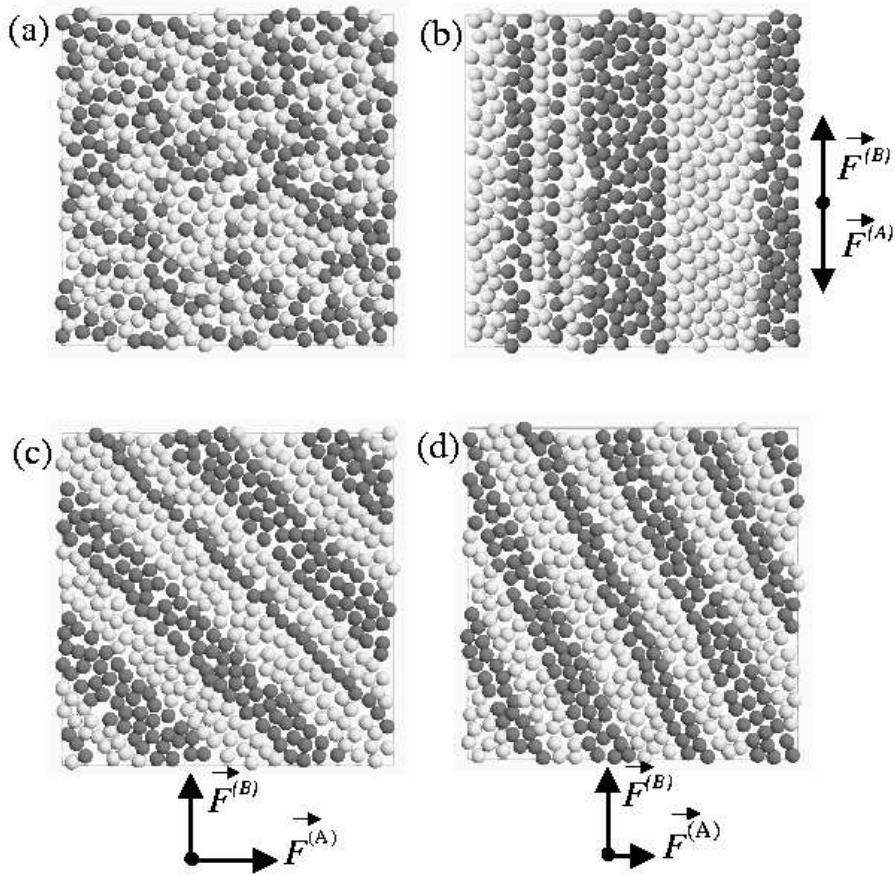, width=12cm, angle=0}
    \caption{Typical simulation snapshots of the two dimensional system: (a)
      disordered state without field, (b) lane formation
      with parallel fields $\vec F^{(A)}=-\vec F^{(B)}=180 k_{B}T/\sigma$ in
      $y$-direction above the critical force. (c) lane 
      formation with perpendicular  fields of same magnitude
      $|\vec F^{(A)}|=|\vec F^{(B)}|=180 k_{B}T/\sigma$  above the  critical force
      difference. (d) lane formation  with  perpendicular fields and  $|\vec
      F^{(A)}|/|\vec F^{(B)}|=1/2$ with $|\vec F^{(B)}|=180 k_{B}T/\sigma $ above the
      critical force difference. 
      In (c) and (d) the $A$  particles are drifting in 
      $x$-direction, while $B$ particles are 
      drifting in $y$-direction. The lanes are moving
      perpendicular to $\Delta \vec F$. The particles
      are  depicted as 
      spheres with diameter $\sigma$. A light sphere is an
      $A$-particle while a  gray sphere is a $B$-particle.}    
  \label{ss_2d}
\end{center}
\end{figure}

In our simulation we assume - without loss of generality -
the direction of ${\vec F}^{(B)}$ along the $y$ axis of the simulation box.
The model parameters are fixed to $\rho\sigma^{2}=1.0$,
$\kappa=4.0$, and $V_{0}=2.5 k_{B}T$.
Simulation snapshots for different external field strengths are shown in
Fig. \ref{ss_2d}(a)-(d).

In Fig. \ref{ss_2d}(a), no field is applied and a homogeneous
completely mixed state is visible. In Fig. \ref{ss_2d}(b),
on the other hand, the external forces are parallel: $\vec
F^{(A)}=-\vec F^{(B)}$. 
The magnitude $|\vec F^{(A)}-\vec F^{(B)}| $ is beyond the
critical strength \cite{jo} such that lane formation parallel to the field
shows up.
Nonparallel forces with perpendicular directions are
investigated in  Figs. \ref{ss_2d} (c) and (d). 
One observes formation of tilted lanes which are indeed
in the direction of the force difference vector as expected from
our general argument. In following the configurations as a function of time we
verified the simple formula of the interface velocity $v_I$ as given
in Eq.\  (\ref{velocity}).

\begin{figure}
\begin{center}
    \epsfig{file=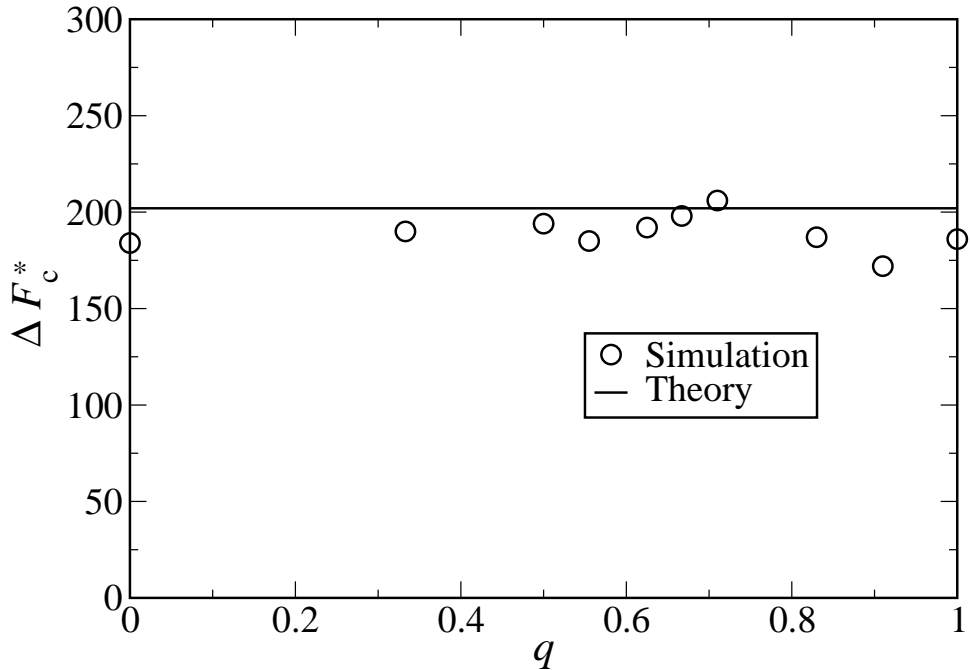, width=11cm, angle=-90}
    \caption{Dimensionless critical force difference $\Delta  F^{*}_{\rm c}=\Delta
      F_{\rm c}\sigma/k_{B}T$ versus the
      ratio $q$ between the perpendicular forces $|\vec F^{(A)}|$ and
      $|\vec F^{(B)|} $ for the two-dimensional system. Above the critical force
      the system is in the patterned state characterized by stripe
      formation. The circles are simulation results, while the solid
      line is theory. The parameters are $\kappa=4.0$,
      $V_{0}=2.5 k_{B}T$, and $\rho\sigma^{2}=1.0$. }  
  \label{phase}
\end{center}
\end{figure}

In a next step, we study perpendicular forces keeping  their
ratio $q=| {\vec F}^{(A)}| /| {\vec F}^{(B)}|$
fixed but increasing their magnitude.
For small forces the system stays demixed. In increasing the
strength of both fields, we have calculated suitable order parameters 
which are sensitive to tilted lane formation. These are immediate
generalizations of those used in Ref. \cite{jo}. The order parameter
exhibits a sharp jump indicating the critical field strength
of the force difference vector
upon which tilted lane formation is achieved. There is a clear hysteresis
loop if the force is reduced again
such that the non-equilibrium phase transition towards tilted lane
formation is of first order.
The  critical value $\Delta F_{c}$ of the force difference
$| \Delta {\vec F} |$ 
is shown versus the ratio 
$q=| {\vec F}^{(A)}| /| {\vec F}^{(B)}|$
in Fig. \ref{phase}. We shall compare these data to a simple 
theoretical prediction in the next subsection.

\begin{figure}
\begin{center}
    \epsfig{file=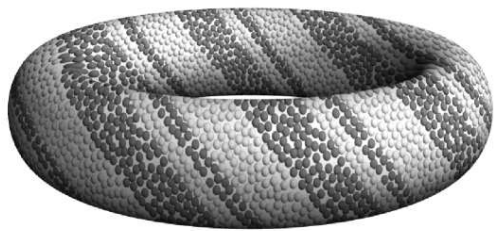, width=7cm, angle=0} 
    \caption{Visualization of the two-dimensional system by mapping
      a two-dimensional simulation snapshot onto the surface of a
      torus. $A$ particles
      are drifting along the torus, while $B$ particles are drifting
      around it. The lanes are moving  perpendicular to their
      direction.}        
  \label{torus}
\end{center}
\end{figure}

A careful remark is in order for small positive $q$. The periodic boundary
conditions used in the simulation corresponds to a toroidal topology
shown in Fig. \ref{torus}. If $q$ is getting small the boundary conditions enforce
a multiple winding around the torus such that finite size effects are expected to be 
significant. Therefore we have not shown simulation data for small $q$
in Fig. \ref{ss_2d}.

\subsection{Theory}
We are aiming at a rough theoretical estimation of the
boundaries of the demixing transition with constant external
fields $\vec F^{(A)}$ and $\vec F^{(B)}$. This was already
put forward in Ref.\ \cite{jo} for parallel forces and is generalized
here to the general case of non-parallel forces. 
\begin{figure}
\begin{center}
    \epsfig{file=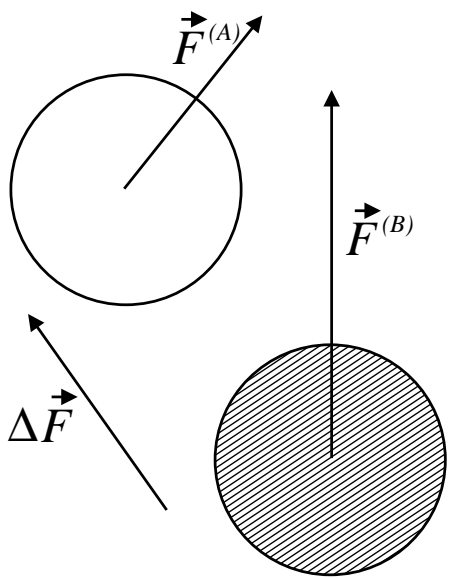, width=6cm, angle=0}
    \caption{Sketch of two colloids of opposite type colliding due to the
      external fields $\vec F^{(A)}$ and $\vec F^{(B)}$. The  $A$ particle is
      white, while the $B$ particle is gray. The particles collide
      effectively in direction of the difference force $\Delta \vec
      F=\vec F^{(B)}-\vec F^{(A)}$ as seen from a fixed
      center-of-mass.}  
  \label{sketch}
\end{center}
\end{figure}
Consider first a central collision between an $A$ and $B$ particle
 pair, see Fig. \ref{sketch}. Transforming the trajectories onto one
 with a fixed common center of mass of the two particles, one realizes
 that the collision is effectively driven by half of the force difference,
 $\frac{1}{2}(\vec F^{(A)}-\vec F^{(B)})=-\frac{1}{2}\Delta\vec F$ for
 $A$ particles and $\frac{1}{2}(\vec F^{(B)}-\vec F^{(A)})=\frac{1}{2}\Delta\vec F$ for
 $B$ particles. A transition towards patterned lanes is expected if
 $|\Delta \vec F|/2$ is larger than a typical  {\it average force} 
 between $A$ and $B$ particles, and  lane formation is induced. The
 latter  force  
depends both on density and on the external fields themselves. We
estimate a typical average force 
between two opposite particles by considering different ``effective''
interparticle spacings. The first typical interparticle spacing is set
by the density alone, $a=\rho^{-1/2}$. Including fluctuations in the
interparticle distance induced by a finite temperature results in a
further smaller effective average distance $\tilde a$ as obtained by
setting a typical interparticle energy equal to $V(a)+k_{B}T$. Hence
$\tilde a=V^{-1}\left[V(a)+k_{B}T\right]$ where $V^{-1}$ is the
inverse function of the interaction potential $V(r)$. Finally the
presence of the external fields enforces an even smaller averaged
distance $a'$ between two colliding opposite particles. 

 We estimate this minimum distance $a'$ by adding the net
force per colliding particle $\Delta F/2$ to the force at distance
$\tilde a$ via
\begin{eqnarray}
a'=F^{-1}\left[\Delta F/2+F(\tilde a)\right ],
\label{c1}
\end{eqnarray}
where $F^{-1}$ is the inverse function of $F(r)=-\frac{d}{dr} V(r)$. In
general, an $AB$ particle pair will not collide directly along
$\Delta \vec F/2$ such
that the actual average distance is between $a'$ and $\tilde a$. Hence
the averaged force $\overline f$ between an $A$ and a $B$
particle is roughly 
\begin{eqnarray}
{\overline f_{\rm }}=\frac{1}{\tilde a-a'}[V(a')-V(\tilde a)].
\label{c2}
\end{eqnarray}
The critical force difference $\Delta F_{\rm c}$ is reached when it
 becomes of the order of the mean force
$\overline f$,
\begin{eqnarray}
\Delta F_{\rm c} = 2 \lambda \overline f.
\label{c3}
\end{eqnarray}
$\lambda=2$ is a dimensionless prefactor which
is determined by an optimal fit to all simulation results
for parallel forces \cite{jo}.

The basic prediction of this simple theory is that the only essential
parameter governing tilted lane formation is the magnitude
 $| {\Delta \vec F} |$. This prediction can be tested by simulation.
In fact, in Fig. \ref{phase}, the critical value of $|{\Delta \vec F} |$
is shown for different ratios $q=| {\vec F}^{(A)}| /| {\vec F}^{(B)}|$.
Were the theory correct, all the simulation data should fall
on a horizontal line independent of $q$. As can be deduced from 
Fig. \ref{phase}, this is indeed confirmed. Furthermore, the actual
magnitude predicted from the theory is in line with the simulation data.
Note that - as far as the non-parallel case is concerned - there is no
fit parameter involved. The global fit parameter $\lambda$ is solely adjusted to
the case of parallel forces.

\subsection{Three-dimensional model}
\begin{figure}
\begin{center}
    \epsfig{file=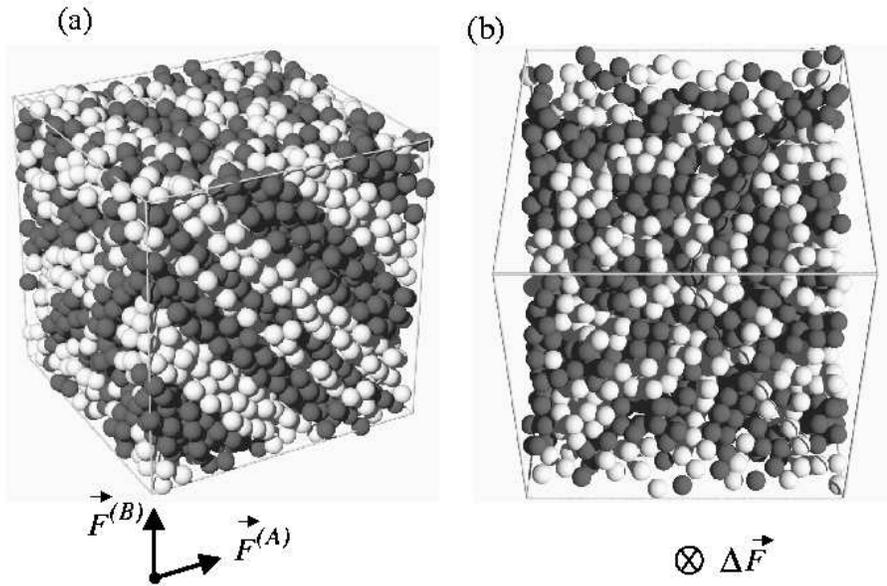, width=12cm, angle=0} 
    \caption{Typical snapshots of the three-dimensional system with
      perpendicular external fields above the critical force
      difference. The magnitude of the forces is $|\vec F^{(A)}|=|\vec
      F^{(B)}|=150 k_{B}T/\sigma$. (a) Three-dimensional view, (b) look 
      on the plane perpendicular to the $\Delta \vec F$ vector.
cd} 
  \label{ss_3d}
\end{center}
\end{figure}

The model and all methods and arguments can readily
be generalized to three spatial
dimensions. Similar conclusions hold for the formation of
tilted lanes. We have also performed computer simulations in a cubic
box in three dimensions and observed tilted lane formation. Results
are presented in Fig. \ref{ss_3d}: tilted lane formation is clearly
visible in the plane spanned by the two forces $\vec F^{(A)}$ and
$\vec F^{(B)}$, see Fig. \ref{ss_3d} (a). 
Perpendicular to the direction $ \Delta {\vec F}$ of the lanes,
the system shows a structure reminiscent
of two-dimensional spinodal decomposition, see Fig. \ref{ss_3d} (b).
The parameters are for these snapshots  $\kappa=4.0$, $V_{0}=2.5
k_{B}T$ and $\rho\sigma^{3}=1$. In conclusion, this shows that pattern
formation is a general effect which is independent of the
dimensionality of the model.
\section{Reentrant freezing for a driven Brownian crystal}
\label{reentrant}
In this section we focus on a Brownian {\it crystal}
which is driven by an external field. To this end,
the external fields acting onto $A$ and $B$ particles are parallel 
(${\vec F}^{(A)}=-{\vec F}^{(B)}=f{\vec e}_{y}$)
but in contrast to Ref. \cite{jo} the field-free equilibrium initial state is a 
triangular crystal. The case of a fluid field-free  configuration
is easier as rotational symmetry with respect to the direction of the external
field applied is ensured. This is no longer true for a crystal where
one has to specify the field direction with respect to the crystalline orientation
resulting in an anisotropy which reflects the crystalline symmetry.
Clearly, in the absence of any external field, $A$ and $B$ particles are indistinguishable;
hence the equilibrium state is randomly occupied triangular crystal.
In the other limit of very strong fields, one expects again phase separation
into completely demixed $A$ and $B$ regions. Once they are demixed, they
follow Boltzmann statistics. Consequently the equilibrium state is a
pure $A$ (or $B$) crystal of the same lattice than the original (field
free) one. What is less clear intuitively is how the system
transforms from the first randomly occupied crystal into the demixed crystal
if the field is turned continuously on. At least two scenarios
are conceivable:
either the system retains the underlying solid lattice but particle exchange hopping
processes generated by the external field demix the crystalline state or
the crystal first melts mechanically via the external field and then crystallizes again.
In our simulations we almost exclusively observed the latter scenario.

In order to detect a triangular crystalline order we define a suitable crystallinity
order parameter  $\Psi_{6}$ that
probes sixfold symmetry around a given
particle via   \cite{Heni,Heni:pattern}
    \begin{eqnarray}
      \label{psi_6}
      \Psi_{6} = \left|\left<\frac{1}{12N}
          \sum_{j=1}^{2N}\sum_{<k>} e^{6 {\rm i}
            \phi_{jk}}\right>\right|.
\label{psi}
    \end{eqnarray}
Here the $k$-sum includes the six nearest neighbors of the given particle
and the $j$-sum extends over $2N$ particles in the simulation box. 
The large angular brackets indicate a time average. $\phi_{jk}$ is the
polar angle of the interparticle distance 
vector with respect to a fixed reference frame. For ideal sixfold
symmetry, i.e., for a perfect triangular crystal, $\Psi_{6}=1$. 
Thermal fluctuations cause  deviations from this ideal case but
a value of  $\Psi_{6}>0.8$
\cite{Heni,Heni:pattern} is conveniently taken to be conclusive evidence for a triangular crystal.

\begin{figure}
\begin{center}
    \epsfig{file=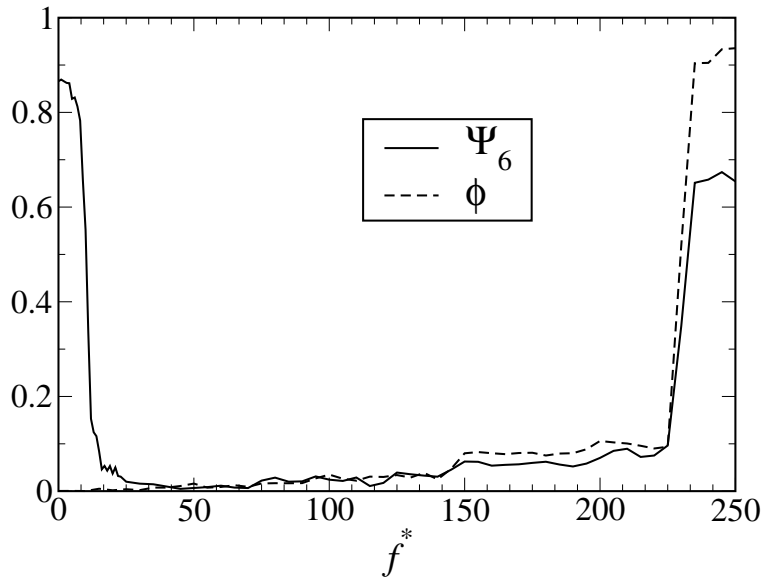, width=9cm, angle=-90} 
    \caption{Crystallinity order parameter $\Psi_{6}$ (solid line)
      and lane order parameter $\phi$ (dashed line)
      plotted versus reduced external force $f^{*}$ as calculated by a
      Brownian dynamics simulation.}     
  \label{order parameter}
\end{center}
\end{figure}

\begin{figure}
\begin{center}
    \epsfig{file=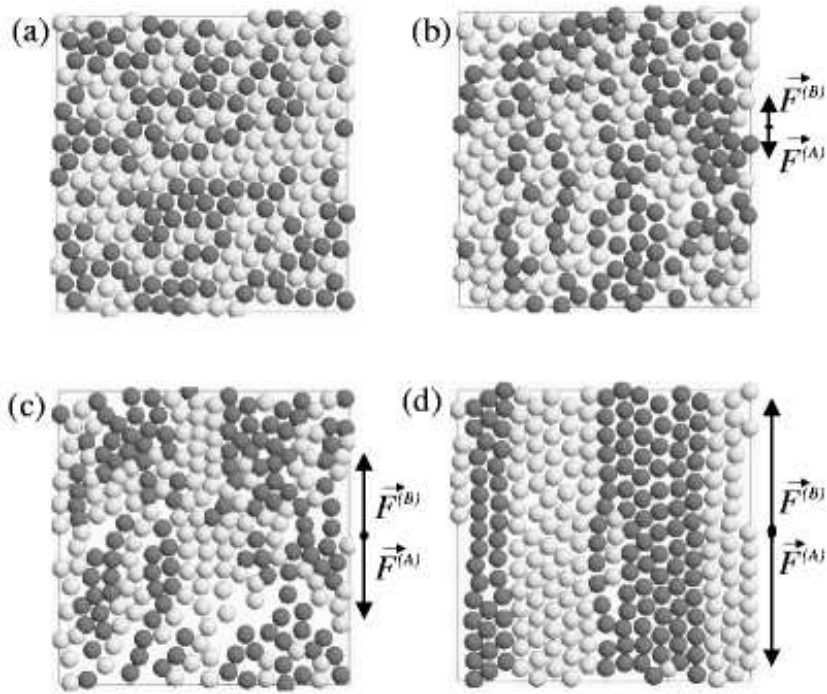, width=11cm, angle=0} 
    \caption{Snapshots of the two-dimensional system for different
      external forces, starting with a solid in the free-field
      state. The forces acting on the two different particle 
      types are pointing into opposite directions. The forces are
      (a) $f^{*}=0$, (b) $f^{*}=50$, (c) $f^{*}=150$, (d)
      $f^{*}=250$. In (a) and (d), the system shows a solid structure, while
      in (b) and (c) the system is a homogeneously fluid. Here the total
      number of particles is $2N=250$.
}    
  \label{sssolid}
\end{center}
\end{figure}

In Fig. \ref{order parameter} we plot $\Psi_{6}$ versus the difference
external
force $f^{*}=|\vec F^{(B)}|\sigma/k_{B}T$ for fixed parameters $\kappa=4.0, V_{0}=15.0 k_{B}T$, and
$\rho\sigma^{2}=1$. 
The direction
of the force
is (11)-direction of the triangular crystal. 
Note that in contrast to the parameters
used in section \ref{nonparallel}, the interactions energy is much
larger to ensure
that the equilibrium field-free state is crystalline.
Indeed the crystallinity order parameter in the field-free case
is around  $\Psi_{6}=0.87$. Upon increasing the external
field strength to $f^{*}\approx 10$, the crystallinity order
parameter sharply drops down and stays to values close to zero
indicating a melting of the crystal induced by the external field.
This  melting process  is mainly caused by a mechanical stress 
induced by the external field with respect to $AB$ particle pairs.
As the field is getting larger ($f^{*} \gtrsim 200$), the order
parameter $\Psi_{6}$ increases again
to values close to 0.7. This is accompanied with lane formation
as indicated by a drastic increase of another order
parameter $\phi$ defined in Ref.\ \cite{jo} which is sensitive to lane formation.
The whole scenario is illustrated also by simulation  snapshots
 shown in Fig. \ref{sssolid}. While Fig. \ref{sssolid} (a)
corresponds to a field-free randomly mixed crystal, Fig. \ref{sssolid} (b)
and (c) are in the molten state while Fig. \ref{sssolid} (d) represents
a refrozen demixed crystal sliding against each other. In
Fig. \ref{sssolid} (b) and (c) worm like structures along the fields
occur as a precursor to lanes formed by solids.
Consequently we have shown evidence for a {\it reentrant freezing}
behavior generated by external fields in non-equilibrium. A qualitative
similar situation occurs for  colloidal solids in linear shear flow
\cite{AvB5,AvB10,Onuki}.
A continuous increase of the shear rate  can lead to shear-melting
and subsequent recrystallization into a different solid structure
\cite{Stevens,Lahiri,Lahiri2}. A similar effect is shear thinning
and subsequent shear thickening as observed in colloidal fluids for
increasing shear rates \cite{newstein}. 

\section{Conclusions}
\label{conclusions}
To summarize: we have generalized earlier studies of non-equilibrium pattern formation
in continuum driven diffusive mixtures  to non-parallel external fields
and crystalline states. As main results we found  tilted lanes along
the force difference vector which are wandering with a constant
interface velocity provided the external force difference is large enough.
Furthermore, a solid melts and refreezes if the magnitude
of an external field is increased.

Let us remark on possibilities of non-equilibrium lane formation
in more complicated systems: first, our system studied was completely symmetric
involving the same partial densities and the same particle-particle interactions.
For experimental realizations
\cite{weiland,batchelor,yan:1993,nasreldin:1999} this will not be
fulfilled in general. However, the basic physics of lane formation
will not change. Secondly, if ternary and further 
multicomponent mixture beyond binary ones are considered, we expect cascades
of lane formation transitions involving the different particles
species as the external field is increased.

We finally comment on possible experimental realizations of our model:
There are different fields where the pattern formation we predicted
within our model can be verified, namely in colloidal dynamics and in
pedestrian motion.  {\it Binary colloidal mixtures} indeed can be
driven by constant external forces. Important examples for
{\underline {parallel forces}} are sedimentation where the external
force is gravity \cite{weiland,batchelor} or electrokinetic motion of charged
colloids \cite{mantegazza} where the external force is an electric
field. Both the fluid and crystalline field-free state can be studied.
A recent realization with mixed crystals can be found in
\cite{wette}. One 
possible drawback is the hydrodynamic backflow \cite{Weitz} caused by
strong hydrodynamic interactions \cite{Gerhard,Dhontbook,ladd:1993} which were
neglected in our model. An overall backflow can be avoided by a time
dependent oscillatory  field (e.g. AC electric field) which leads
qualitatively to the same lane formation if its frequency is small
enough \cite{jo}. We further think that the long-ranged  hydrodynamic
flow around a driven colloidal particle will favor lane formation,
i.e.\ the critical field strength needed to generate lane formation is
expected to be lower than with hydrodynamic interactions neglected.
Colloids can also be exposed to external laser-optical and magnetic
filed \cite{top} which generate external forces in a controlled way
coupling to the dielectricity (resp.\ the magnetic permeability) of the
colloidal material.
{\underline {Nonparallel external forces}} in colloidal mixtures can be
realized by crossing two external fields e.g. gravity with electric,
laser-optical with electric, laser-optical with magnetic etc. The two
species of a colloidal mixture will in general respond differently to
the two external fields such that the resulting total external force
will be different in direction. 

A different realization is {\it dynamics of pedestrians} in pedestrian
zones and in lecture halls. Similar off-lattice models involving
Brownian particles have been used to simulate the collective behavior
of pedestrians \cite{Helbing,Helbing_neu} including escape panic
\cite{Helbing:panic}. Our setup of perpendicular external fields is
realized by two crossing pedestrian lanes in which pedestrians are
only moving in one direction. Based on our results, we would expect
tilted lane formation provided the density of the pedestrians is high
enough. Finally it would be interesting to extend phenomenological
hydrodynamical theories which predict lane formation for parallel
forces via an instability
\cite{kynch:1952,batchelor,valiveti:1998,valiveti:1999,burger:2000,biesheuvel:2001,jay}
to the case of tilted forces. Work along these lines is in progress.  

\section*{Acknowledgments} 
It is with great pleasure that we dedicate this paper to
J. P. Hansen on the occasion of his 60th birthday.
We  thank T. Palberg, J. Chakrabarti, J. Sherwood, V. Popkov and
E. Allahyarov for helpful remarks. Financial support from the DFG
(Sonderforschungsbereich 237) is gratefully acknowledged.


\section*{References}

\begin{thebibliography}{10}
\expandafter\ifx\csname bibnamefont\endcsname\relax
  \def\bibnamefont#1{#1}\fi
\expandafter\ifx\csname bibfnamefont\endcsname\relax
  \def\bibfnamefont#1{#1}\fi
\expandafter\ifx\csname url\endcsname\relax
  \def\url#1{\texttt{#1}}\fi
\expandafter\ifx\csname urlprefix\endcsname\relax\def\urlprefix{URL }\fi
\expandafter\ifx\csname bibinfo\endcsname\relax \def\bibinfo#1#2{#2}\fi
\expandafter\ifx\csname eprint\endcsname\relax \def\eprint#1{#1}\fi

\bibitem{pattern}
\bibinfo{author}{\bibfnamefont{M.~C.} \bibnamefont{Cross}} \bibnamefont{and}
  \bibinfo{author}{\bibfnamefont{P.~C.} \bibnamefont{Hohenberg}},
  \bibinfo{journal}{Rev. Mod. Phys.} \textbf{\bibinfo{volume}{65}},
  \bibinfo{pages}{851} (\bibinfo{year}{1993}).

\bibitem{Saarloos}
\bibinfo{author}{\bibfnamefont{W.}~\bibnamefont{van Saarloos}},
  \bibinfo{journal}{Phys. Rep.} \textbf{\bibinfo{volume}{301}},
  \bibinfo{pages}{9} (\bibinfo{year}{1998}).

\bibitem{Allen}
\bibinfo{author}{\bibfnamefont{M.~P.} \bibnamefont{Allen}} \bibnamefont{and}
  \bibinfo{author}{\bibfnamefont{D.~J.} \bibnamefont{Tildesley}},
  \emph{\bibinfo{title}{Computer Simulations of Liquids}}
  (\bibinfo{publisher}{Clarendon Press}, \bibinfo{address}{Oxford},
  \bibinfo{year}{1989}).

\bibitem{review1}
\bibinfo{editor}{\bibfnamefont{M.}~\bibnamefont{Baus}},
  \bibinfo{editor}{\bibfnamefont{L.}~\bibnamefont{Rull}}, \bibnamefont{and}
  \bibinfo{editor}{\bibfnamefont{J.}~\bibnamefont{Ryckaert}}, eds.,
  \emph{\bibinfo{title}{Observations, Prediction and Simulation of Phase
  Transitions in Complex Fluids}}, vol. \bibinfo{volume}{Series B: Physics}
  (\bibinfo{publisher}{Kluwer Academic Publishers},
  \bibinfo{address}{Dordrecht}, \bibinfo{year}{1995}).

\bibitem{Hartmutr}
\bibinfo{author}{\bibfnamefont{H.}~\bibnamefont{L{\"o}wen}},
  \bibinfo{journal}{Phys. Rep.} \textbf{\bibinfo{volume}{237}},
  \bibinfo{pages}{249} (\bibinfo{year}{1994}).

\bibitem{HansenMcDonald}
\bibinfo{author}{\bibfnamefont{J.~P.} \bibnamefont{Hansen}} \bibnamefont{and}
  \bibinfo{author}{\bibfnamefont{I.~R.} \bibnamefont{McDonald}},
  \emph{\bibinfo{title}{Theory of Simple Liquids}}
  (\bibinfo{publisher}{Academic Press}, \bibinfo{address}{London},
  \bibinfo{year}{1986}), 2 ed.

\bibitem{Ziareview}
\bibinfo{author}{\bibfnamefont{B.}~\bibnamefont{Schmittmann}} \bibnamefont{and}
  \bibinfo{author}{\bibfnamefont{R.~K.~P.} \bibnamefont{Zia}},
  \emph{\bibinfo{title}{{\rm in} Phase Transitions and Critical Phenomena}},
  vol.~\bibinfo{volume}{17} (\bibinfo{publisher}{Academic Press},
  \bibinfo{address}{London}, \bibinfo{year}{1995}), \bibinfo{note}{ed. by C.
  Domb and J. Lebowitz}.

\bibitem{ramaswamy1}
\bibinfo{author}{\bibfnamefont{S.}~\bibnamefont{Ramaswamy}},
  \bibinfo{journal}{Current Science} \textbf{\bibinfo{volume}{77}},
  \bibinfo{pages}{402} (\bibinfo{year}{1999}).

\bibitem{colloid_non}
\bibinfo{author}{\bibfnamefont{V.}~\bibnamefont{Trappe}},
  \bibinfo{author}{\bibfnamefont{V.}~\bibnamefont{Prasad}},
  \bibinfo{author}{\bibfnamefont{L.}~\bibnamefont{Cipelletti}},
  \bibinfo{author}{\bibfnamefont{P.~N.} \bibnamefont{Segre}}, \bibnamefont{and}
  \bibinfo{author}{\bibfnamefont{D.~A.} \bibnamefont{Weitz}},
  \bibinfo{journal}{Nature} \textbf{\bibinfo{volume}{411}},
  \bibinfo{pages}{772} (\bibinfo{year}{2001}).

\bibitem{top}
\bibinfo{author}{\bibfnamefont{H.}~\bibnamefont{L{\"o}wen}},
  \bibinfo{journal}{J. Phys.: Condens. Matter} \textbf{\bibinfo{volume}{13}},
  \bibinfo{pages}{R 415} (\bibinfo{year}{2001}).

\bibitem{Murray1996}
\bibinfo{author}{\bibfnamefont{C.~A.} \bibnamefont{Murray}} \bibnamefont{and}
  \bibinfo{author}{\bibfnamefont{D.~G.} \bibnamefont{Grier}},
  \bibinfo{journal}{Annu. Rev. Phys. Chem.} \textbf{\bibinfo{volume}{47}},
  \bibinfo{pages}{421} (\bibinfo{year}{1996}).

\bibitem{Helbing}
\bibinfo{author}{\bibfnamefont{D.}~\bibnamefont{Helbing}},
  \bibinfo{author}{\bibfnamefont{I.~J.} \bibnamefont{Farkas}},
  \bibnamefont{and} \bibinfo{author}{\bibfnamefont{T.}~\bibnamefont{Vicsek}},
  \bibinfo{journal}{Phys. Rev. Lett.} \textbf{\bibinfo{volume}{84}},
  \bibinfo{pages}{1240} (\bibinfo{year}{2000}).

\bibitem{Helbing_neu}
\bibinfo{author}{\bibfnamefont{D.}~\bibnamefont{Helbing}},
  \bibinfo{author}{\bibfnamefont{P.}~\bibnamefont{Moln\'{a}r}},
  \bibinfo{author}{\bibfnamefont{I.~J.} \bibnamefont{Farkas}},
  \bibnamefont{and} \bibinfo{author}{\bibfnamefont{K.}~\bibnamefont{Bolay}},
  \bibinfo{journal}{Environment and Planning B: Planning and Design}
  \textbf{\bibinfo{volume}{28}}, \bibinfo{pages}{361} (\bibinfo{year}{2001}).

\bibitem{schadschneider}
\bibinfo{author}{\bibfnamefont{C.}~\bibnamefont{Burstedde}},
  \bibinfo{author}{\bibfnamefont{K.}~\bibnamefont{Klauck}},
  \bibinfo{author}{\bibfnamefont{A.}~\bibnamefont{Schadschneider}},
  \bibnamefont{and} \bibinfo{author}{\bibfnamefont{J.}~\bibnamefont{Zittartz}},
  \bibinfo{journal}{Physica A} \textbf{\bibinfo{volume}{295}},
  \bibinfo{pages}{507} (\bibinfo{year}{2001}).

\bibitem{Hoffmann1}
\bibinfo{author}{\bibfnamefont{G.~P.} \bibnamefont{Hoffmann}} \bibnamefont{and}
  \bibinfo{author}{\bibfnamefont{H.}~\bibnamefont{L{\"o}wen}},
  \bibinfo{journal}{Phys. Rev. E} \textbf{\bibinfo{volume}{60}},
  \bibinfo{pages}{3009} (\bibinfo{year}{1999}).

\bibitem{Hoffmann3}
\bibinfo{author}{\bibfnamefont{G.~P.} \bibnamefont{Hoffmann}} \bibnamefont{and}
  \bibinfo{author}{\bibfnamefont{H.}~\bibnamefont{L{\"o}wen}},
  \bibinfo{journal}{J. Phys.: Condens. Matter} \textbf{\bibinfo{volume}{13}},
  \bibinfo{pages}{9197} (\bibinfo{year}{2001}).

\bibitem{jo}
\bibinfo{author}{\bibfnamefont{J.}~\bibnamefont{Dzubiella}},
  \bibinfo{author}{\bibfnamefont{G.~P.} \bibnamefont{Hoffmann}},
  \bibnamefont{and}
  \bibinfo{author}{\bibfnamefont{H.}~\bibnamefont{L{\"o}wen}},
  \bibinfo{journal}{Phys. Rev. E} \textbf{\bibinfo{volume}{65}}
  (\bibinfo{year}{2002}), \bibinfo{note}{in press}.

\bibitem{rothman:1994}
\bibinfo{author}{\bibfnamefont{D.~H.} \bibnamefont{Rothman}} \bibnamefont{and}
  \bibinfo{author}{\bibfnamefont{S.}~\bibnamefont{Zaleski}},
  \bibinfo{journal}{Reviews of Modern Physics} \textbf{\bibinfo{volume}{66}},
  \bibinfo{pages}{1417} (\bibinfo{year}{1994}).

\bibitem{weiland}
\bibinfo{author}{\bibfnamefont{R.~H.} \bibnamefont{Weiland}},
  \bibinfo{author}{\bibfnamefont{Y.~P.} \bibnamefont{Fessas}},
  \bibnamefont{and} \bibinfo{author}{\bibfnamefont{B.~V.}
  \bibnamefont{Ramaro}}, \bibinfo{journal}{J. Fluid. Mech.}
  \textbf{\bibinfo{volume}{142}}, \bibinfo{pages}{383} (\bibinfo{year}{1984}).

\bibitem{batchelor}
\bibinfo{author}{\bibfnamefont{G.~K.} \bibnamefont{Batchelor}}
  \bibnamefont{and} \bibinfo{author}{\bibfnamefont{R.~W.~J.} \bibnamefont{van
  Rensburg}}, \bibinfo{journal}{J. Fluid. Mech.}
  \textbf{\bibinfo{volume}{166}}, \bibinfo{pages}{379} (\bibinfo{year}{1986}).

\bibitem{nasreldin:1999}
\bibinfo{author}{\bibfnamefont{H.~A.} \bibnamefont{Nasr-El-Din}},
  \bibinfo{author}{\bibfnamefont{J.~H.} \bibnamefont{Masliyah}},
  \bibnamefont{and}
  \bibinfo{author}{\bibfnamefont{K.}~\bibnamefont{Nandakumar}},
  \bibinfo{journal}{Canadian Journal of Chemical Engineering}
  \textbf{\bibinfo{volume}{11}}, \bibinfo{pages}{1003} (\bibinfo{year}{1999}).

\bibitem{yan:1993}
\bibinfo{author}{\bibfnamefont{Y.}~\bibnamefont{Yan}} \bibnamefont{and}
  \bibinfo{author}{\bibfnamefont{J.~H.} \bibnamefont{Masliyah}},
  \bibinfo{journal}{International Journal of Multiphase Flow}
  \textbf{\bibinfo{volume}{19}}, \bibinfo{pages}{875} (\bibinfo{year}{1993}).

\bibitem{Hone}
\bibinfo{author}{\bibfnamefont{E.}~\bibnamefont{Chang}} \bibnamefont{and}
  \bibinfo{author}{\bibfnamefont{D.~W.} \bibnamefont{Hone}},
  \bibinfo{journal}{Europhys. Lett.} \textbf{\bibinfo{volume}{5}},
  \bibinfo{pages}{635} (\bibinfo{year}{1988}).

\bibitem{Loewen92}
\bibinfo{author}{\bibfnamefont{H.}~\bibnamefont{L{\"o}wen}},
  \bibinfo{journal}{J. Phys.: Condens. Matter} \textbf{\bibinfo{volume}{4}},
  \bibinfo{pages}{10105} (\bibinfo{year}{1992}).

\bibitem{Loehle}
\bibinfo{author}{\bibfnamefont{B.}~\bibnamefont{L{\"o}hle}} \bibnamefont{and}
  \bibinfo{author}{\bibfnamefont{R.}~\bibnamefont{Klein}},
  \bibinfo{journal}{Physica A} \textbf{\bibinfo{volume}{235}},
  \bibinfo{pages}{224} (\bibinfo{year}{1997}).

\bibitem{Naidoo}
\bibinfo{author}{\bibfnamefont{K.~J.} \bibnamefont{Naidoo}} \bibnamefont{and}
  \bibinfo{author}{\bibfnamefont{J.}~\bibnamefont{Schnitker}},
  \bibinfo{journal}{J. Chem. Phys.} \textbf{\bibinfo{volume}{100}},
  \bibinfo{pages}{3114} (\bibinfo{year}{1994}).

\bibitem{Hoffmann2}
\bibinfo{author}{\bibfnamefont{G.~P.} \bibnamefont{Hoffmann}} \bibnamefont{and}
  \bibinfo{author}{\bibfnamefont{H.}~\bibnamefont{L{\"o}wen}},
  \bibinfo{journal}{J. Phys.: Condens. Matter} \textbf{\bibinfo{volume}{12}},
  \bibinfo{pages}{7359} (\bibinfo{year}{2000}).

\bibitem{Roux}
\bibinfo{author}{\bibfnamefont{H.}~\bibnamefont{L{\"o}wen}},
  \bibinfo{author}{\bibfnamefont{J.~P.} \bibnamefont{Hansen}},
  \bibnamefont{and} \bibinfo{author}{\bibfnamefont{J.~N.} \bibnamefont{Roux}},
  \bibinfo{journal}{Phys.\ Rev.\ A} \textbf{\bibinfo{volume}{44}},
  \bibinfo{pages}{1169} (\bibinfo{year}{1991}).

\bibitem{Ermak}
\bibinfo{author}{\bibfnamefont{D.~L.} \bibnamefont{Ermak}},
  \bibinfo{journal}{J. Chem. Phys} \textbf{\bibinfo{volume}{62}},
  \bibinfo{pages}{4189} (\bibinfo{year}{1975}).

\bibitem{santra:1996}
\bibinfo{author}{\bibfnamefont{S.~B.} \bibnamefont{Santra}},
  \bibinfo{author}{\bibfnamefont{S.}~\bibnamefont{Schwarzer}},
  \bibnamefont{and} \bibinfo{author}{\bibfnamefont{H.}~\bibnamefont{Herrmann}},
  \bibinfo{journal}{Phys. Rev. E} \textbf{\bibinfo{volume}{54}},
  \bibinfo{pages}{5066} (\bibinfo{year}{1996}).

\bibitem{Heni}
\bibinfo{author}{\bibfnamefont{M.}~\bibnamefont{Heni}} \bibnamefont{and}
  \bibinfo{author}{\bibfnamefont{H.}~\bibnamefont{L{\"o}wen}},
  \bibinfo{journal}{Phys. Rev. Lett.} \textbf{\bibinfo{volume}{85}},
  \bibinfo{pages}{3668} (\bibinfo{year}{2000}).

\bibitem{Heni:pattern}
\bibinfo{author}{\bibfnamefont{M.}~\bibnamefont{Heni}} \bibnamefont{and}
  \bibinfo{author}{\bibfnamefont{H.}~\bibnamefont{L{\"o}wen}},
  \bibinfo{journal}{J. Phys.: Condens. Matter} \textbf{\bibinfo{volume}{13}},
  \bibinfo{pages}{4675} (\bibinfo{year}{2001}).

\bibitem{AvB5}
\bibinfo{author}{\bibfnamefont{A.}~\bibnamefont{Imhof}},
  \bibinfo{author}{\bibfnamefont{A.}~\bibnamefont{van Blaaderen}},
  \bibnamefont{and} \bibinfo{author}{\bibfnamefont{J.~K.~G.}
  \bibnamefont{Dhont}}, \bibinfo{journal}{Langmuir}
  \textbf{\bibinfo{volume}{10}}, \bibinfo{pages}{3477} (\bibinfo{year}{1994}).

\bibitem{AvB10}
\bibinfo{author}{\bibfnamefont{T.}~\bibnamefont{Palberg}} \bibnamefont{and}
  \bibinfo{author}{\bibfnamefont{M.}~\bibnamefont{W{\"u}rth}},
  \bibinfo{journal}{J. Phys. I (France)} \textbf{\bibinfo{volume}{6}},
  \bibinfo{pages}{237} (\bibinfo{year}{1996}).

\bibitem{Onuki}
\bibinfo{author}{\bibfnamefont{A.}~\bibnamefont{Onuki}}, \bibinfo{journal}{J.
  Phys.: Condens. Matter} \textbf{\bibinfo{volume}{9}}, \bibinfo{pages}{6119}
  (\bibinfo{year}{1997}).

\bibitem{Stevens}
\bibinfo{author}{\bibfnamefont{M.~J.} \bibnamefont{Stevens}} \bibnamefont{and}
  \bibinfo{author}{\bibfnamefont{M.~O.} \bibnamefont{Robbins}},
  \bibinfo{journal}{J. Chem. Phys.} \textbf{\bibinfo{volume}{98}},
  \bibinfo{pages}{2319} (\bibinfo{year}{1993}).

\bibitem{Lahiri}
\bibinfo{author}{\bibfnamefont{R.}~\bibnamefont{Lahiri}} \bibnamefont{and}
  \bibinfo{author}{\bibfnamefont{S.}~\bibnamefont{Ramaswamy}},
  \bibinfo{journal}{Phys. Rev. Lett.} \textbf{\bibinfo{volume}{73}},
  \bibinfo{pages}{1043} (\bibinfo{year}{1994}).

\bibitem{Lahiri2}
\bibinfo{author}{\bibfnamefont{R.}~\bibnamefont{Lahiri}} \bibnamefont{and}
  \bibinfo{author}{\bibfnamefont{S.}~\bibnamefont{Ramaswamy}},
  \bibinfo{journal}{Physica A} \textbf{\bibinfo{volume}{224}},
  \bibinfo{pages}{84} (\bibinfo{year}{1996}).

\bibitem{newstein}
\bibinfo{author}{\bibfnamefont{M.~C.} \bibnamefont{Newstein}},
  \bibinfo{author}{\bibfnamefont{H.}~\bibnamefont{Wang}},
  \bibinfo{author}{\bibfnamefont{N.~P.} \bibnamefont{Balsara}},
  \bibinfo{author}{\bibfnamefont{A.~A.} \bibnamefont{Lefebvre}},
  \bibinfo{author}{\bibfnamefont{Y.}~\bibnamefont{Shnidman}},
  \bibinfo{author}{\bibfnamefont{H.}~\bibnamefont{Watanabe}},
  \bibinfo{author}{\bibfnamefont{K.}~\bibnamefont{Osaki}},
  \bibinfo{author}{\bibfnamefont{T.}~\bibnamefont{Shikita}},
  \bibinfo{author}{\bibfnamefont{H.}~\bibnamefont{Niwa}}, \bibnamefont{and}
  \bibinfo{author}{\bibfnamefont{Y.}~\bibnamefont{Morishima}},
  \bibinfo{journal}{J. Chem. Phys.} \textbf{\bibinfo{volume}{111}},
  \bibinfo{pages}{4827} (\bibinfo{year}{1999}).

\bibitem{mantegazza}
\bibinfo{author}{\bibfnamefont{F.}~\bibnamefont{Mantegazza}},
  \bibinfo{author}{\bibfnamefont{V.}~\bibnamefont{Degiorgio}},
  \bibinfo{author}{\bibfnamefont{A.~V.} \bibnamefont{Delgado}},
  \bibnamefont{and} \bibinfo{author}{\bibfnamefont{F.~J.}
  \bibnamefont{Arroyo}}, \bibinfo{journal}{J. Chem. Phys.}
  \textbf{\bibinfo{volume}{109}}, \bibinfo{pages}{6905} (\bibinfo{year}{1998}).

\bibitem{wette}
\bibinfo{author}{\bibfnamefont{P.}~\bibnamefont{Wette}},
  \bibinfo{author}{\bibfnamefont{H.~J.} \bibnamefont{Schope}},
  \bibinfo{author}{\bibfnamefont{R.}~\bibnamefont{Biehl}}, \bibnamefont{and}
  \bibinfo{author}{\bibfnamefont{T.}~\bibnamefont{Palberg}},
  \bibinfo{journal}{J. Chem. Phys.} \textbf{\bibinfo{volume}{114}},
  \bibinfo{pages}{7556} (\bibinfo{year}{2001}).

\bibitem{Weitz}
\bibinfo{author}{\bibfnamefont{P.~N.} \bibnamefont{Segre}},
  \bibinfo{author}{\bibfnamefont{F.}~\bibnamefont{Liu}},
  \bibinfo{author}{\bibfnamefont{P.}~\bibnamefont{Umbanhowar}},
  \bibnamefont{and} \bibinfo{author}{\bibfnamefont{D.~A.} \bibnamefont{Weitz}},
  \bibinfo{journal}{Nature} \textbf{\bibinfo{volume}{409}},
  \bibinfo{pages}{594} (\bibinfo{year}{2001}).

\bibitem{Gerhard}
\bibinfo{author}{\bibfnamefont{G.}~\bibnamefont{N{\"a}gele}},
  \bibinfo{journal}{Phys. Rep.} \textbf{\bibinfo{volume}{272}},
  \bibinfo{pages}{215} (\bibinfo{year}{1996}).

\bibitem{Dhontbook}
\bibinfo{author}{\bibfnamefont{J.~K.~G.} \bibnamefont{Dhont}},
  \emph{\bibinfo{title}{An Introduction to Dynamics of Colloids}}
  (\bibinfo{publisher}{Elsevier}, \bibinfo{address}{Amsterdam},
  \bibinfo{year}{1996}).

\bibitem{ladd:1993}
\bibinfo{author}{\bibfnamefont{A.~J.~C.} \bibnamefont{Ladd}},
  \bibinfo{journal}{Physics of Fluids A} \textbf{\bibinfo{volume}{5}},
  \bibinfo{pages}{299} (\bibinfo{year}{1993}).

\bibitem{Helbing:panic}
\bibinfo{author}{\bibfnamefont{D.}~\bibnamefont{Helbing}},
  \bibinfo{author}{\bibfnamefont{I.~J.} \bibnamefont{Farkas}},
  \bibnamefont{and} \bibinfo{author}{\bibfnamefont{T.}~\bibnamefont{Vicsek}},
  \bibinfo{journal}{Nature} \textbf{\bibinfo{volume}{407}},
  \bibinfo{pages}{487} (\bibinfo{year}{2000}).

\bibitem{kynch:1952}
\bibinfo{author}{\bibfnamefont{G.~J.} \bibnamefont{Kynch}},
  \bibinfo{journal}{Trans. Faraday Soc.} \textbf{\bibinfo{volume}{48}},
  \bibinfo{pages}{166} (\bibinfo{year}{1952}).

\bibitem{valiveti:1998}
\bibinfo{author}{\bibfnamefont{P.}~\bibnamefont{Valiveti}} \bibnamefont{and}
  \bibinfo{author}{\bibfnamefont{D.~L.} \bibnamefont{Koch}},
  \bibinfo{journal}{Applied Scientific Research} \textbf{\bibinfo{volume}{58}},
  \bibinfo{pages}{275} (\bibinfo{year}{1998}).

\bibitem{valiveti:1999}
\bibinfo{author}{\bibfnamefont{P.}~\bibnamefont{Valiveti}} \bibnamefont{and}
  \bibinfo{author}{\bibfnamefont{D.~L.} \bibnamefont{Koch}},
  \bibinfo{journal}{Physics of Fluids} \textbf{\bibinfo{volume}{11}},
  \bibinfo{pages}{3283} (\bibinfo{year}{1999}).

\bibitem{burger:2000}
\bibinfo{author}{\bibfnamefont{R.}~\bibnamefont{Burger}},
  \bibinfo{author}{\bibfnamefont{F.}~\bibnamefont{Concha}},
  \bibinfo{author}{\bibfnamefont{K.~K.} \bibnamefont{Fjelde}},
  \bibnamefont{and} \bibinfo{author}{\bibfnamefont{K.~H.}
  \bibnamefont{Karlsen}}, \bibinfo{journal}{Powder Technology}
  \textbf{\bibinfo{volume}{113}}, \bibinfo{pages}{30} (\bibinfo{year}{2000}).

\bibitem{biesheuvel:2001}
\bibinfo{author}{\bibfnamefont{P.~M.} \bibnamefont{Biesheuvel}},
  \bibinfo{author}{\bibfnamefont{H.}~\bibnamefont{Verweij}}, \bibnamefont{and}
  \bibinfo{author}{\bibfnamefont{V.}~\bibnamefont{Breedveld}},
  \bibinfo{journal}{Aiche Journal} \textbf{\bibinfo{volume}{47}},
  \bibinfo{pages}{45} (\bibinfo{year}{2001}).

\bibitem{jay}
\bibinfo{author}{\bibfnamefont{J.}~\bibnamefont{Chakrabarti}},
  \bibinfo{author}{\bibfnamefont{J.}~\bibnamefont{Dzubiella}},
  \bibnamefont{and} \bibinfo{author}{\bibfnamefont{H.}~\bibnamefont{L{\"o}wen}}
   (\bibinfo{year}{2002}), \bibinfo{note}{to be published}.

\end{thebibliography}
\end{document}